\documentclass[sigconf]{acmart}

\AtBeginDocument{%
  \providecommand\BibTeX{{%
    \normalfont B\kern-0.5em{\scshape i\kern-0.25em b}\kern-0.8em\TeX}}}

\copyrightyear{2022}
\acmYear{2022}
\setcopyright{rightsretained}
\acmConference[EASE 2022]{The International Conference on Evaluation and Assessment in Software Engineering 2022}{June 13--15, 2022}{Gothenburg, Sweden}
\acmBooktitle{The International Conference on Evaluation and Assessment in Software Engineering 2022 (EASE 2022), June 13--15, 2022, Gothenburg, Sweden}

\usepackage{tabularx}
\begin{document}

\title{Personality Traits in Game Development}

\author{Miriam Sturdee}
\affiliation{%
 \institution{Lancaster University}
 \city{Lancaster}
 \country{United Kingdom}}
\email{m.sturdee@lancaster.ac.uk}

\author{Matthew Ivory}
\affiliation{%
 \institution{Lancaster University}
 \city{Lancaster}
 \country{United Kingdom}}
\email{matthew.ivory@lancaster.ac.uk}

\author{David Ellis}
\affiliation{%
 \institution{University of Bath}
 \city{Bath}
 \country{United Kingdom}}
\email{dae30@bath.ac.uk}

\author{Patrick Stacey}
\affiliation{%
 \institution{Loughborough University}
 \city{Loughborough}
 \country{United Kingdom}}
\email{P.Stacey@lboro.ac.uk}

\author{Paul Ralph}
\affiliation{%
 \institution{Dalhousie University}
 \city{Halifax}
 \country{Canada}}
\email{paulralph@dal.ca}

\renewcommand{\shortauthors}{Sturdee et al.}

\begin{abstract}
Existing work on personality traits in software development excludes game developers as a discrete group. Whilst games are software, game development has unique considerations, so game developers may exhibit different personality traits from other software professionals. We assessed responses from 123 game developers on an International Personality Item Pool Five Factor Model scale and demographic questionnaire using factor analysis. Programmers reported lower Extraversion than designers, artists and production team members; lower Openness than designers and production, and reported higher Neuroticism than production---potentially linked to burnout and crunch time. Compared to published norms of software developers, game developers reported lower Openness, Conscientiousness, Extraversion and Agreeableness, but higher Neuroticism. These personality differences have many practical implications: differences in Extraversion among roles may precipitate communication breakdowns; differences in Openness may induce conflict between programmers and designers. Understanding the relationship between personality traits and roles can help recruiters steer new employees into appropriate roles, and help managers apply appropriate stress management techniques. To realise these benefits, individuals must be distinguished from roles: just because an individual occupies a role does not mean they possess personality traits associated with that role.

\end{abstract}



\maketitle

\section{Introduction}
Game development is a large segment within software development, with approximately 2.5 billion players and global turnover of US\$180.1 billion in 2021.\footnote{ \url{https://techjury.net/blog/video-game-demographics/}}. While video games are software systems, they are qualitatively different from other kinds of software systems. Banking software, for example, differs from video games in both purpose and intended audience. 

Game development entails different (if not unique) challenges including its `crunch-time' culture: when big releases are imminent, developers are required to work around the clock to ensure deadlines are met~\cite{edholm2017crunch}. Game development also entails more heterogeneous roles than many other kinds of software development, including graphic design, sound design, voice acting, 2D and 3D art, story writing, physics, as well as programming for multiple platforms and unique types of testing (e.g. play-testing). 

Multidisciplinary game development teams attract more diverse employees~\cite{feldtLinksPersonalitiesViews2010}. While quantifying such diversity is difficult, there may be something specific about the games industry that inspires people beyond other software roles (e.g. working with financial software). Understanding why people gravitate to these roles allows the industry to adapt and evolve, include people displaying relevant traits, or recruit others into the various roles inherent in this multi-dimensional industry. 

One way to explore the background of game developers, and identify proficiency in skill sets of certain roles within game development, is through personality traits~\cite{capretzCallPromoteSoft2018}. Studying personality traits can reveal factors underlying outlook and behaviour, and produce insights into how a person's personality evolves over their lifespan~\cite{costa2019personality}. Personality scales are often used by employers to estimate a person's suitability for a role, but can also be used in other contexts, or for individual curiosity. 
	
Existing studies of software development and personality have either studied broad populations~\cite{cruzFortyYearsResearch2015}, examined particular application areas such as software testing~\cite{kanij2015empirical}, or looked at cultural differences~\cite{aminImpactPersonalityTraits2020a, akarsuManagingSocialAspects2020}, yet few studies explicitly examine personality traits for game developers, though some examine personality traits of \textit{players}~\cite{nacke2014brainhex, potard2020video}. We aim to extend the existing body of knowledge for personality in software development by exploring the personality traits of workers in the games development industry. We therefore propose the following research question: \textit{Do game developers have strong indicators of traits on any particular personality axes, and do the personality traits of game developers differ from those of software developers in general?}

Due to the novelty of the population, we conduct an exploratory study to identify the personality traits of game developers. We define \textit{personality} in the context of psychological research into traits; that is, a set of behavioural characteristics, preferences, and patterns that are relatively stable within an individual's lifespan~\cite{roberts2000rank}, many of which make up a person's `personality'. We define a \textit{game developer} as a person engaged in creating or maintaining a video game including programming, design, management and production, with a focus on the technical aspects of the field.

Next, we outline psychology research on personality traits, including the Five Factor Model (FFM) and Myers Briggs Type Indicator (MBTI\textsuperscript{\textregistered}), before explaining how this research applies to software engineering and games development. Then, in Section \ref{sec:Method}, we describe our instrumentation, recruitment strategy, hypotheses, and analytical approach. Section \ref{sec:ResultsAnalysis} includes data diagnostics, descriptive statistics, factor analysis and group comparisons. We then discuss how the results may be applicable to the real world, compare possibilities in games versus software development, and consider the next steps and implications for this novel path of research in Section \ref{sec:Discussion}. Section \ref{sec:Conclusion} concludes the paper with a summary of its contributions. 

Given the apparent lack of related research in this area, and its potential, we aim to produce an initial investigation into personality traits in game development, in order to support ongoing work on personality in the wider field of software development.

\section{Related Work}
\label{sec:RelatedWork}
A common view in psychology is that personalities are a collection of behavioural traits, preferences and patterns~\cite{matthewsPersonalityTraits2003}, which are relatively stable over time~\cite{mccraeStabilityPersonalityObservations1994}. Subsequently, personality traits can explain and predict behaviours. Predicting and understanding behaviours exhibited by particular personality traits facilitates identifying specific traits that may draw people to certain professions or careers. Research on personality traits of software engineers goes back to the early 1990s~\cite{ezoeAssessmentPersonalityTraits1994}, likely associated with the rise of the internet and increased levels of software within businesses. 

Personality is mostly measured using standardised questionnaires. Answers are combined using factor analysis to estimate quantitatively an individual's personality traits (factors). These questions are carefully curated and presented to limit biases. Several notable scales exist but the scales used most often in software engineering research are the International Personality Item Pool (IPIP) items~\cite{goldbergInternationalPersonalityItem2006}, which includes scales for dozens of personality traits including the Big Five Model (see Section \ref{sec:FFM}), and the Myers-Briggs Type Indicator\textsuperscript{\textregistered} (MBTI) test~\cite{myersMyersBriggsTypeIndicator1962} (see Section \ref{sec:MBTI}). 

\subsection{The Five Factor Model}
\label{sec:FFM}
The Five Factor Model (FFM)~\cite{mccrae1992introduction} proposes five dimensions of human personality: 
\begin{itemize}
    \item Openness to Experience (e.g. amenable to embracing new activities or unconventional ideas)
    \item Conscientiousness (e.g. self-discipline, reliability, dependability)
    \item Extraversion (e.g. exhibiting social confidence, outgoing) 
    \item Agreeableness (e.g. cooperation, friendliness, altruism), 
    \item Neuroticism (e.g. nervousness, moodiness, emotional instability)
\end{itemize}

Each trait has six facets~\cite[cf.][]{kajonius2019assessing} but analysis of individual facets is not necessary for the purposes of this paper. FFM can be measured using the items from the International Personality Item Pool (IPIP)~\cite{goldbergInternationalPersonalityItem2006}, specifically, IPIP-50 (50-item scale) and IPIP-20 (20-item scale)~\cite{goldbergInternationalPersonalityItem2006}. FFM and its associated IPIP scales have been validated in large scale studies with thousands of respondents~\cite[e.g.][]{twomey2021open}.

FFM has formed the theoretical basis of several studies of software professionals and has been extensively empirically tested and validated since its inception in 2006~\cite{goldbergInternationalPersonalityItem2006, gomez2014development}. Within the software professional population, Feldt et al.~\cite{feldtLinksPersonalitiesViews2010} investigated the personalities of 47 professional software engineers using the IPIP-50 scale. They found two personality clusters within their sample, of moderate and intense personality types (those scoring higher in Openness and Extraversion). Those who reported a more intense personality were typically younger, had a preference for multi-tasking and for contributing to a project as opposed to working with it from beginning to the end. Kosti et al.~\cite{kostiPersonalityEmotionalIntelligence2014} found similar results in computer science postgraduate students using the IPIP-20 scale, particularly of moderate and intense style personalities. Those with intense personalities were more likely to prefer working in teams and management-style roles within the development process. Despite using different versions of the IPIP, the studies found comparable results which suggests reliability within the IPIP.
	
Calefato et al.~\cite{calefatoDevelopersPersonalityLargeScale2018} used a different method---text mining across web pages of the Apache Software Foundation, mailing list archives and git repositories---to examine the differences in developer personalities. The authors found that those who reported higher Agreeableness and Openness were more likely to have authored more successfully-integrated commits into a git repository. In terms of recruitment activities, we might suggest that individuals exhibiting those particular traits could appeal to employers. The authors also found that as developers evolve during their careers, they become more neurotic, conscientious and agreeable, which suggests that personality traits are changeable, with situations and contexts influencing traits~\cite{heckman2012hard}, which indicates that traits are not stable over the lifetime of an individual~\cite{mottus2012personality}. 

In contrast, Amin et al.~\cite{aminTraitBasedPersonalityProfile2018} investigated personality profiles of programmers in Pakistan and found that Agreeableness and Conscientiousness were the most dominant traits, with Neuroticism being the least dominant trait. Extraversion was also relatively low. Meanwhile, in a Turkish sample, Extraversion was very low while Conscientiousness was high~\cite{akarsuManagingSocialAspects2020}. Cross-cultural factors may therefore influence personality in software development. In a different vein, Doungsa-ard et al.~\cite{doungsa-ardSoftwareEngineeringPosition2020} used a mapping analysis to  sort new software engineers automatically into the most appropriate role for their personality based upon data from current occupiers of similar roles, highlighting the distinctions between roles, such as analyst, designer, developer and tester. Despite an increasing number of studies in recent years however, only a few datasets using IPIP scales for the software development population exist~\cite{goldbergInternationalPersonalityItem2006}. This hinders modeling software engineer personality types based solely upon IPIP data.

\subsection{Myers-Briggs Type Indicator}
\label{sec:MBTI}

Many studies of personality in software engineering and elsewhere use MBTI~\cite{barrosoInfluenceHumanPersonality2017, cruzFortyYearsResearch2015}, which classifies personalities on four dimensions that combine to form a person's `preferred' personality type~\cite{myersMyersBriggsTypeIndicator1962}. These four dimensions are Extraversion-Introversion, Sensing-Intuition, Thinking-Feeling and Judging-Per\-ceiving. A meta-analysis indicates the two most common personality types in software development are ISTJ and INTJ~\cite{barrosoInfluenceHumanPersonality2017}, suggesting that most programmers are introverts who rely on logic to make decisions and prefer more structure in their working environment, whereas `feeling' types were much less well represented. Research based on MBTI\textsuperscript{\textregistered} should be interpreted with caution. MBTI\textsuperscript{\textregistered} is a commercial product based on Jung's unvalidated psychoanalysis theory~\cite{jung1916contribution}. Despite its prominence in hiring strategies, much research disputes its validity and usefulness ~\cite[e.g.][]{boyleMyersBriggsTypeIndicator2009, stromberg2015myers} and argues for better models of personality~\cite{mccraeReinterpretingMyersBriggsType1989}, including FFM.

\subsection{Personality and Job Roles}
Software engineering includes diverse job roles and specialisms~\cite{capretzCallPromoteSoft2018}, far beyond the stereotypical programmer~\cite{capretzInfluencePersonalityTypes2015}. For example, a contemporary application development team might comprise a product manager, product designer, several developers, and specialists in quality assurance and DevOps. Previous work has highlighted the most dominant personalities within specific roles, (see Section \ref{sec:RelatedWork}), to help people choose career paths. Prior research into role separation has exclusively relied on MBTI\textsuperscript{\textregistered}.
	
One such study of 100 software engineers cross-referenced the roles of analyst, designer, programmer, tester and maintainer against MBTI\textsuperscript{\textregistered} personality types. They found that 84\% of extroverts preferred analyst roles compared to 70\% of introverts preferring design-based roles and 65\% of introverts preferring programming~\cite{capretzInfluencePersonalityTypes2015}. Similar results were found by Ahmed et al. ~\cite{ahmedEvaluatingDemandSoft2012} and Capretz et al.\cite{capretzMakingSenseSoftware2010}. Capretz et al.~\cite{capretzWhyWeNeed2010} also mapped the skills of software roles to personalities within particular positions. They found that EF types were well suited to analysis roles, NT types for design, IST for programming, SJ for testing and SP for maintenance roles, which is line with previous findings~\cite{ahmedEvaluatingDemandSoft2012, capretzMakingSenseSoftware2010, capretzInfluencePersonalityTypes2015}. Similarly, Hardiman et al.~\cite{hardimanPersonalityTypesSoftware1997} suggested that the best programmers were typically NT and SJ types, but software engineering has evolved greatly since the publication of this paper in~\citeyear{hardimanPersonalityTypesSoftware1997}. Finally, in comparison to MBTI\textsuperscript{\textregistered} studies, Acuna et al.~\cite{acunaEmphasizingHumanCapabilities2006} found that a team leader should display certain traits, and that these traits differ from the traits best for programming or testing. Acuna et al.~\cite{acunaHowPersonalityTeam2009} later found more Agreeable, Conscientious students report higher job satisfaction.

\subsection{Applicability to Game Development}
Whilst existing software engineering research has investigated the intersection of personality and role, no one has yet investigated the personality types present within the game development community. Game developers may differ from other software developers because games encompass a wider range of creative and artistic tasks~\cite{smutsAreVideoGames2005}. Game development may therefore attract people who have strong creative traits alongside those who enjoy programming. While we found no studies of personality in game developers, Amin et al.~\cite{aminImpactPersonalityTraits2020a} found that programmer creativity is directly related to Openness and Extraversion and inversely related to Neuroticism. This makes intuitive sense as more open, extraverted people tend to seek out new experiences, which may foster original ideas that can be channeled into their creative work. 

\section{Method}
\label{sec:Method}
We developed and administered a questionnaire survey to collect personality, demographic, and employment data from game development professionals over a one-month period. 

\subsection{Hypotheses}

We begin with two hypotheses: 

\noindent \textbf{H1:} Game developers in different roles will report distinct \-personality traits

\noindent \textbf{H2:} Game developers will report distinct personality traits compared to those previously reported by software developers

Both of these hypotheses follow from the basic tenant of personality research that certain tasks, roles or jobs tend to appeal more to people with certain personality traits. Consequently, people tend to perform better at jobs that align with their personality traits. As described in Section \ref{sec:RelatedWork}, previous research found personality differences among roles within software development organizations. Analogously, we expect to find personality differences both within game development organizations and between game development and other kinds of software development.

\subsection{Instrumentation}
We implemented a questionnaire survey using \url{Qualtrics(.com)}. The study obtained ethical approval from the host institution, and informed consent was given prior to starting the survey. The questionnaire began with a published, standardised introduction to personality testing in which participants were asked to describe themselves honestly as they generally were at that moment in time. We operationalized the Five Factor Model using the IPIP-50 scale~\cite{buchananImplementingFiveFactorPersonality2005}, with five-point rating items ranging from ``strongly agree'' to ``strongly disagree''. IPIP questions were presented in random order. Associated positive/negative scoring for each factor was programmed into the questionnaire to allow participants to see their trait tendencies after taking part. Participants also answered basic demographic questions about themselves and their employment, and were asked to select their role(s) (e.g. programming) and sub-roles (e.g. Engine Programmer). Open text boxes were provided for non-standard answers.

\subsection{Recruitment}
The questionnaire was administered both online and in-person (using tablets configured to record answers in a fresh questionnaire). Participants either self-selected via a `walk-up' at an English-speaking, UK-based industry and trade conference (\textit{Develop:Brighton}), or completed the same materials online via links posted on Twitter, either by responding to the advertisement directly, or via snowball sampling. The online component of the study ran in parallel to the conference data collection and the study also remained open for four weeks after the conference ended. Participants were not compensated for their participation, but were able to view their trait scores at the end of the survey.

\subsection{Analysis Procedure}

Data were analysed in R (version 4.1.0). We used Exploratory Factor Analysis (EFA) to assess convergent and discriminant validity, and remove problematic items. We then used Confirmatory Factor Analysis (CFA) to re-affirm correct loadings and compute trait (factor) scores. Finally, we used one-way ANOVA to test our hypotheses by assessing personality trait differences. Post-hoc Tukey tests and eta-squared effect sizes were conducted for significant ANOVA results, with a \emph{p}-value of $<.05$ as the boundary for significance. One-way ANOVA comparisons are made against two published norms of software developers~\cite{feldtLinksPersonalitiesViews2010, kostiPersonalityEmotionalIntelligence2014} to evaluate whether game developers display significantly different personality profiles than the broader umbrella term of software developer.

\section{Results}
\label{sec:ResultsAnalysis}
This section explores personality differences within game development teams and between our sample of game developers and previously-studied samples of developers of other kinds of software.

\subsection{Descriptive Statistics}

Data from 128 participants was considered usable, but five participants were excluded from the analysis due to belonging to very small roles: Audio and Quality Assurance (1 and 4 participants, respectively). Given the unique skills and experience required for these roles, it was more appropriate to remove them than to re-categorise them. As such, 123 participants were included in the analysis. Of the 123 participants, 26 identified as female and 97 as male. Table \ref{tab:age} provides further demographic information on the reported roles and ages of respondents. 

\begin{table}[ht]
\centering
  \caption{Breakdown of the different roles by age}
  \label{tab:age}
  \begin{tabular}{crrrr|r}
  \toprule
    Age & Art & Design & Production & Programming & Total\\
    \midrule
    16--20 & 0 & 0 & 0 & 1 & 1\\
    21--25 & 6 & 4 & 2 & 10 & 22\\
    26--30 & 3 & 14 & 6 & 15 & 38\\
    31--35 & 7 & 5 & 3 & 12 & 27\\
    36--40 & 2 & 3 & 0 & 10 & 15\\
    41--45 & 2 & 1 & 3 & 4 & 10\\
    46--50 & 1 & 3 & 1 & 1 & 6\\
    51--55 & 0 & 0 & 1 & 0 & 1\\
    56+ & 0 & 2 & 1 & 0 & 3\\
    \midrule
    Total & 21 & 32 & 17 & 53 & \textbf{123}\\
  \bottomrule
\end{tabular}
\end{table}

\subsection{Diagnostics}
To assess complete-incomplete response bias, chi-square comparisons were made for role, $\chi^2(5, 181) = 5.30, \emph{p} = .381$, and gender, $\chi^2(1, 186) < .001, \emph{p} = 1$. Since both tests fail to reject the null hypothesis (that the groups are the same); we find no evidence of response bias. 

\subsection{Scale Validation and Construct Validity}

To assess data factorability prior to factor analysis we applied Bartlett's test of Sphericity and the  Kaiser-Meyer-Olkin (KMO) test for sampling adequacy. Bartlett's test was significant ($p < .001$) indicating relationships between personality items. The KMO score of .68 indicates that the data is acceptable, if not ideal, for factor analysis.

Next, we conducted exploratory factor analysis (EFA) with the weighted least squares estimator and varimax rotation to assess convergent and discriminant validity. We iteratated between dropping the most problematic item and re-running the EFA until all remaining items loaded well (i.e. indicating good convergent and discriminant validity). 

Having removed some problematic items using EFA, we then proceeded with confirmatory factor analysis (CFA), with the diagonal least weighted squares estimator, to further affirm the validity of the scale and compute the individual factor scores. Again, some items appeared problematic and we iterated between dropping the most problematic item and re-running the CFA.

The model converged after 73 iterations, with no significant covariance between factors (\emph{p} < .004 for all factor pairs), indicating successful fit (CFI = .930, TLI = .926 and RMSEA = .083). While these indices are not within the thresholds suggested by~\cite{hu1999cutoff}, such thresholds were designed for continuous data~\cite{xia2019rmsea} and meaningful thresholds for ordinal data, as we have in this study, have been established. Without ordinal data thresholds available for CFA, the improvement in fit indices suggests a better fitting model than with all items (CFI = .887, TLI = .881 and RMSEA = .093). Further items could potentially have been removed to increase item loading scores, but it was considered important to retain as many of the items as possible from the original scale.


Table \ref{tab:EFA_Final} shows the final EFA results, while Table \ref{tab:CFA} shows all standardised loadings for retained items. The factor loadings for each intermediate step are available in the supplementary materials (see \nameref{sec:DataAvailability}, below). In the end, we had to drop six of the Openness items as well as Agreeableness items A2 and A10. With the validation of the data complete and troublesome items removed from further analysis, inferential analysis can proceed.

\begin{table} 
  \centering
  \caption{Final exploratory factor analysis*}
  \label{tab:EFA_Final}
  \begin{tabular}{lrrrrr}
  \toprule
Item & Factor 1 & Factor 2 & Factor 3 & Factor 4 & Factor 5\\
\midrule
A01 &  &  &  & 0.40 & \\
A03 &  &  &  & 0.58 & \\
A04 &  &  &  & 0.55 & \\
A05 &  &  &  & 0.48 & \\
A06 &  &  &  & 0.40 & \\
A07 &  &  &  & 0.41 & \\
A08 &  &  &  & 0.48 & \\
A09 &  &  &  & 0.49 & \\
C01 &  &  & 0.50 &  & \\
C02 &  &  & 0.55 &  & \\
C03 &  &  & 0.50 &  & \\
C04 &  &  & 0.50 &  & \\
C05 &  &  & 0.51 &  & \\
C06 &  &  & 0.34 &  & \\
C07 &  &  & 0.64 &  & \\
C08 &  &  & 0.49 &  & \\
C09 &  &  & 0.51 &  & \\
C10 &  &  & 0.46 &  & \\
E01 & 0.77 &  &  &  & \\
E02 & 0.78 &  &  &  & \\
E03 & 0.45 &  &  &  & \\
E04 & 0.69 &  &  &  & \\
E05 & 0.66 &  &  &  & \\
E06 & 0.70 &  &  &  & \\
E07 & 0.67 &  &  &  & \\
E08 & 0.67 &  &  &  & \\
E09 & 0.50 &  &  &  & \\
E10 & 0.57 &  &  &  & \\
N01 &  & 0.56 &  &  & \\
N02 &  & 0.45 &  &  & \\
N03 &  & 0.65 &  &  & \\
N04 &  & 0.53 &  &  & \\
N05 &  & 0.39 &  &  & \\
N06 &  & 0.57 &  &  & \\
N07 &  & 0.61 &  &  & \\
N08 &  & 0.78 &  &  & \\
N09 &  & 0.80 &  &  & \\
N10 &  & 0.30 &  &  & \\
O03 &  &  &  &  & 0.75\\
O04 &  &  &  &  & 0.40\\
O09 &  &  &  &  & 0.60\\
O10 &  &  &  &  & 0.79\\
\bottomrule
\end{tabular}
\newline
\flushleft Notes: (1) O=Openness, C=Conscientiousness, E=Extraversion, A=Agreeableness, N=Neuroticism; (2) scores <.3 suppressed.
\end{table}

\begin{table}[ht]
  \centering
  \caption{Final confirmatory factor analysis results}
  \label{tab:CFA}
  \begin{tabular}{lrrrrr}
  \toprule
 Item & O & C & E & A & N\\
 \midrule
O03 & 0.76 &  &  &  & \\
O04 & 0.74 &  &  &  & \\
O09 & 0.71 &  &  &  & \\
O10 & 0.88 &  &  &  & \\
C01 &  & 0.48 &  &  & \\
C02 &  & 0.50 &  &  & \\
C03 &  & 0.64 &  &  & \\
C04 &  & 0.70 &  &  & \\
C05 &  & 0.65 &  &  & \\
C06 &  & 0.28 &  &  & \\
C07 &  & 0.73 &  &  & \\
C08 &  & 0.48 &  &  & \\
C09 &  & 0.66 &  &  & \\
C10 &  & 0.63 &  &  & \\
E01 &  &  & 0.76 &  & \\
E02 &  &  & 0.82 &  & \\
E03 &  &  & 0.63 &  & \\
E04 &  &  & 0.67 &  & \\
E05 &  &  & 0.68 &  & \\
E06 &  &  & 0.72 &  & \\
E07 &  &  & 0.77 &  & \\
E08 &  &  & 0.76 &  & \\
E09 &  &  & 0.61 &  & \\
E10 &  &  & 0.70 &  & \\
A01 &  &  &  & 0.72 & \\
A03 &  &  &  & 0.60 & \\
A04 &  &  &  & 0.43 & \\
A05 &  &  &  & 0.54 & \\
A06 &  &  &  & 0.58 & \\
A07 &  &  &  & 0.48 & \\
A08 &  &  &  & 0.64 & \\
A09 &  &  &  & 0.52 & \\
N01 &  &  &  &  & 0.65\\
N02 &  &  &  &  & 0.50\\
N03 &  &  &  &  & 0.75\\
N04 &  &  &  &  & 0.46\\
N05 &  &  &  &  & 0.65\\
N06 &  &  &  &  & 0.68\\
N07 &  &  &  &  & 0.87\\
N08 &  &  &  &  & 0.91\\
N09 &  &  &  &  & 0.88\\
N10 &  &  &  &  & 0.42\\
\bottomrule
\end{tabular}
\newline
\flushleft Notes: (1) O=Openness, C=Conscientiousness, E=Extraversion, A=Agreeableness, N=Neuroticism; (2) scores <.3 suppressed.
\end{table}

\subsection{Role and Personality}    

\subsubsection{Role Comparison.}

\begin{table}[ht]
\centering
  \caption{Difference between roles*}
  \label{tab:anova}
  \begin{tabularx}{\linewidth}{lrrX}
  \toprule
    Trait & $p$ & $\eta^{2}$ & Roles\\
    \midrule
    E & .008 & .09 & Prog.-Design, Prog.-Production\\
    N & .037 & .07 & Prog.-Production\\
    A & .063 & - & no significant differences \\
    C & .110 & - & no significant differences \\
    O & <.001 & .18 & Prog.-Design, Prog.-Art, Prog.-Production\\ 
\bottomrule
\end{tabularx}
\newline
\flushleft *as determined by ANOVA; reported are the \emph{p}-values, the effect size, the roles which differ according to post-hoc Tukey tests
\end{table}

To explore the relationship between personality traits and game development roles, CFA factor scores were compared between roles using one-way ANOVA. One-way ANOVAs were calculated between all roles to compare the significance of role on personality in game developers. Normality was tested with Shapiro-Wilk normality test, with Neuroticism, Agreeableness, Conscientiousness and Extraversion reporting non-significant values. Openness was found to be non-normal with $p=.042$, however one-way ANOVA are found to be robust for non normal data~\cite{blanca2017non}. Due to the relative minor violation of normality for Openness, no further investigation is conducted.

Levene's test for homogeneity was run on each trait, with all returning non-significant values, except for Agreeableness, $p=.046$. Due to the violation, Agreeableness is assessed using Welch's ANOVA (assuming unequal variance).

\begin{figure}[htbp]
  \centering
 \includegraphics[width=\linewidth]{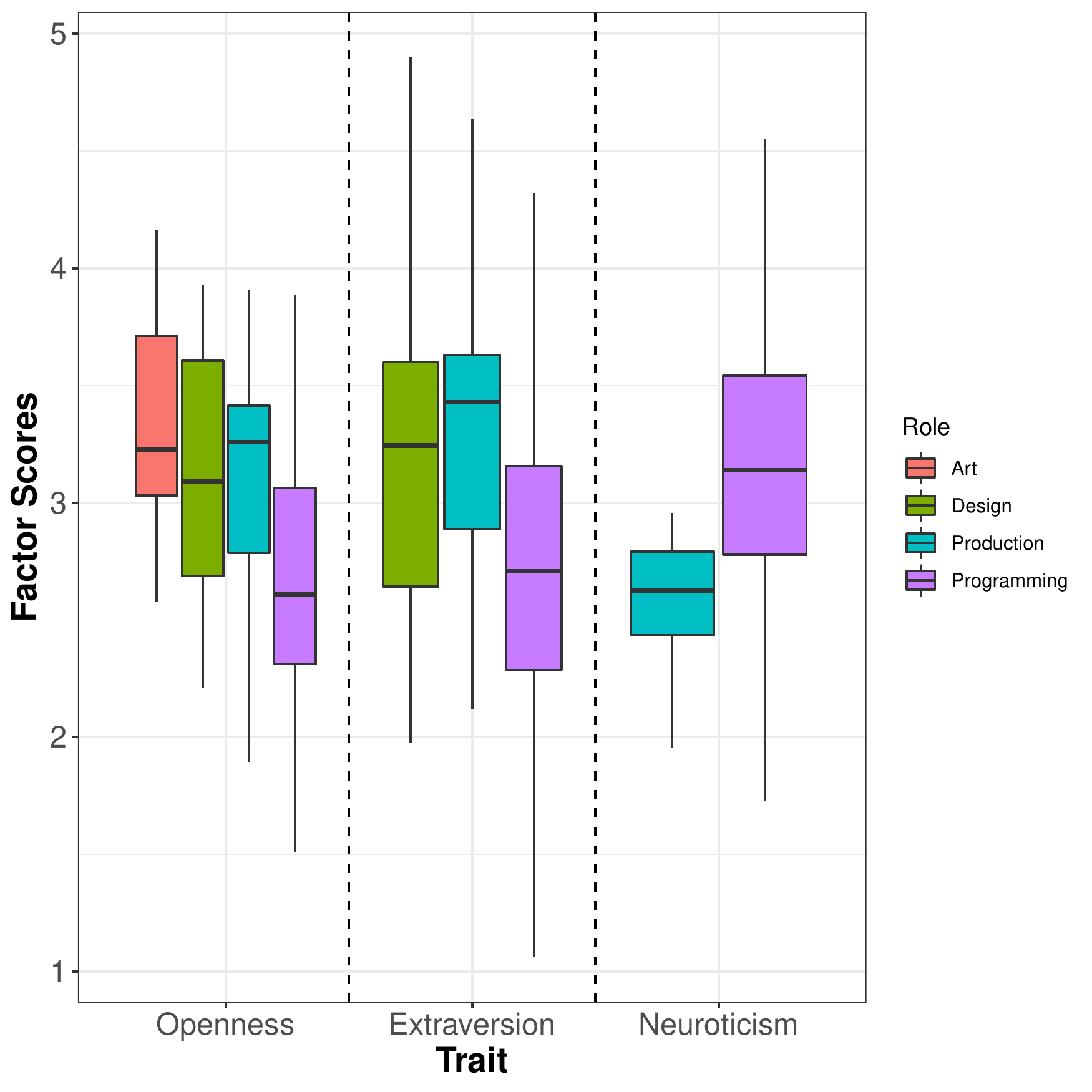}
 \caption{Differences in personality traits between roles (traits with significant differences only)}
   \label{fig:Figure1}
\end{figure}

A significant effect of Openness was seen between roles F(3, 119) = 8.48, \emph{p} < .001, $\eta^{2}$ = .18. A post-hoc Tukey test was conducted to identify which roles were significantly different for Openness and this was found between Programming and Design (adjusted \emph{p} = .002), Programming and Art (adjusted \emph{p} < .001), and Programming and Production (adjusted \emph{p} = .020).

A significant effect for Extraversion was seen F(3, 119) = 4.123, \emph{p} < .008, $\eta^{2}$ = .09. Post-hoc Tukey tests indicated differences between Programming and Design (adjusted \emph{p} = .024), and Programming and Production roles (adjusted \emph{p} = .031).

Comparing Neuroticism between roles with ANOVA identified a significant difference, F(3, 119) = 2.93, \emph{p} < .037. Post-hoc Tukey tests identified differences between the Programming and Production roles (adjusted \emph{p} = .019).

Conscientiousness between roles was not significant, F(3, 119) = 2.06, \emph{p} < .110. 

Agreeableness was analysed using Welch's ANOVA, as this is less sensitive to violation of the homoscedasticity assumption. Agreeableness was not significant, $F(3, 45.2) = 2.61, p < .063$.

Table \ref{tab:anova} summarises these ANOVA results. Based on the substantial inter-role differences in Extraversion, Neuroticism and Openness, \textbf{Hypothesis H1 is supported}.

\subsubsection{Comparison of Game Developers Against Published Norms.}

\begin{figure}[htbp]
  \centering
 \includegraphics[width=\linewidth]{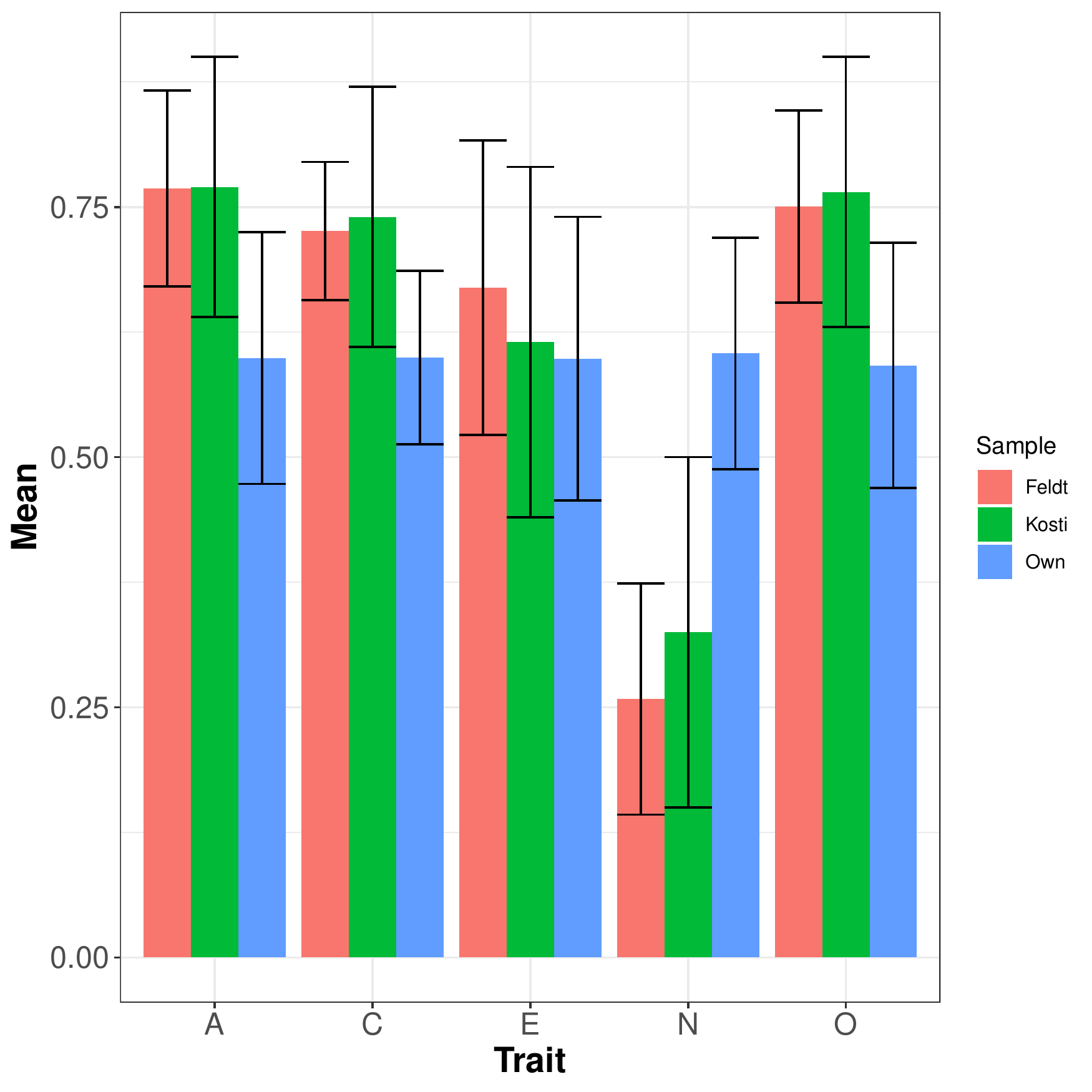}
 \caption{Differences in personality traits between game developers (this study) and software developers~\cite{kostiPersonalityEmotionalIntelligence2014,feldtLinksPersonalitiesViews2010}; means and standard deviations scaled to [0;1]}
   \label{fig:Figure3}
\end{figure}

To explore how game developers' personality profiles differ from typical software developer populations, we compare our results to previously-reported personality profiles~\cite{kostiPersonalityEmotionalIntelligence2014,feldtLinksPersonalitiesViews2010}, comprising 279 software engineering students~\cite{kostiPersonalityEmotionalIntelligence2014} and 47 software engineering professionals~\cite{feldtLinksPersonalitiesViews2010}, respectively. 
The demographics of Feldt et al.'s \cite{feldtLinksPersonalitiesViews2010} study---using software development professionals---align with our sample, with 49\% of their sample defining themselves as programmers, compared to 43\% of ours, with the next largest role in both being design. 

The best way to compare the three studies would be to combine the three original datasets, create a unified factor model (testing for measurement invariance), calculate the factor scores in a consistent manner, and then test for differences using ANOVA. We therefore contacted the authors of the previous studies; however, having been published some years ago, the data was no longer available. 

We therefore proceed by scaling all three data to common range [0,1] to enable comparisons, using ANOVA and post-hoc Tukey tests. (Our analysis scripts are available in supplementary materials---see \nameref{sec:DataAvailability} below).)

Our sample was significantly different from both Feldt et al.'s sample of professionals and Kosti et al.'s sample of software engineering students on all five traits:

\begin{itemize}
\item Openness: F(2, 446) = 80.03, \emph{p} < .001,  $\eta^{2}$ = .26. Post-hoc Tukey tests identified the Game developer sample was significantly lower than both the Kosti sample (adjusted \emph{p} < .001) and the Feldt sample (adjusted \emph{p} < .001).

\item Conscientiousness: F(2, 446) = 65.74, \emph{p} < .001, $\eta^{2}$ = .23. Post-hoc Tukey tests identified the Game developer sample was significantly lower than both the Kosti sample (adjusted \emph{p} < .001) and the Feldt sample (adjusted \emph{p} < .001).

\item Extraversion: F(2, 446) = 3.20, \emph{p} < .042, $\eta^{2}$ = .01. Post-hoc Tukey tests identified the Game developer sample was significantly lower than the Feldt sample (adjusted \emph{p} = .032).

\item Agreeableness: F(2, 446) = 82.23, \emph{p} < .001,  $\eta^{2}$ = .27. Tukey tests indicated that the Game developer sample was significantly lower than both Kosti (adjusted \emph{p} < .001) and Feldt (adjusted \emph{p} < .001).

\item Neuroticism: F(2, 446) = 157.60, \emph{p} < .001,  $\eta^{2}$ = .41. Tukey tests indicated that the Game developer sample was significantly higher than both Kosti (adjusted \emph{p} < .001) and Feldt (adjusted \emph{p} < .001). The Kosti sample was also significantly higher than the Feldt sample (adjusted \emph{p} = .018).
\end{itemize}

Therefore, \textbf{Hypothesis H2 is supported. }

\subsection{Results Summary}
Through analysis of the relationship between roles within game development and personality traits, significant differences were seen with programmers reporting lower Extraversion than Designers, Artists and Production team; lower Openness than Designers and Production, and higher Neuroticism than members of Production. A comparison of the summary statistics of published software developer norms ~\cite{kostiPersonalityEmotionalIntelligence2014,feldtLinksPersonalitiesViews2010} indicated that game developers have lower Agreeableness, Conscientiousness and Openness than both software developer samples; lower Extraversion than professional software developers; and higher Neuroticism than both samples.

\section{Discussion}
\label{sec:Discussion}
The results of this exploratory study indicate a distinct personality profile across game developers as a population, with programmers being a distinct subset within game developers. Game developers present lower levels of Openness, Conscientiousness, Extraversion, and Agreeableness, but higher levels of Neuroticism, than other software developers and software engineering students. Meanwhile, game programmers appear more Neurotic, but less Open and Agreeable than other roles within game development.

A particularly surprising finding was the seemingly high level of Neuroticism in our sample, which may be related to industry and factors such as `crunch time' and burnout~\cite{cote2021cruel, swider2010born}. This could be explained by Edholm et al.'s~\cite{edholm2017crunch} finding that crunch time was predominately a negative experience (e.g. in three out of the four companies studies) and was always associated with higher stress levels. These differences may reflect the competitive nature of the games industry, which may mean those who succeed tend to match the profile and traits we find. 

While it is tempting to assume that game programming attracts more neurotic people, \textit{we cannot discount the possibility that game programming creates stressful environments that cultivates neuroticism.}  Indeed, programmers' higher Neuroticism may be an indication of high levels of job-related stress, exposing employees' neurotic traits more easily than other roles.  More generally, this kind of study cannot distinguish between selection (e.g. introverts are attracted to programming roles), survivorship (e.g. introverts are more likely to stay in programming roles) and causation (e.g. programming increases introversion).

In our sample, game programmers reported lower Extraversion than designers, artists and production team, and lower Openness than designers and production (which might indicate less creativity and originality). Those displaying more introverted behaviours might be better suited to lone working, and working from a provided design specification, rather than generating creative output and getting involved in designing the games. Where individuals work for themselves or in small teams however, all roles are undertaken by the same individuals, which provides an interesting basis for further exploration.  This may be linked to the issue of crunch time in game development, which is likely to affect programmers more than others due to the requirements of the work.

The games industry is inherently creative, as it is concerned with novel entertainment experiences over practical interactions to achieve particular purposes, therefore it may attract developers with a particular interest in creative behaviours and outcomes. Amin et al.'s~\cite{aminTraitBasedPersonalityProfile2018} examination of creativity in programming however suggested that within programmers, creativity was negatively correlated with Neuroticism, meaning this particular group in our cohort at least could be seen as an anomaly in comparison to their participants---or, if we were to speculate, higher Neuroticism may not be correlated with negative impacts on creativity \emph{specifically} in game development overall.

\subsection{Implications}

This study has important implications for game development teams, game development managers and recruiters / career counsellors 

\subsubsection{Implications for Game Development Teams}
Personality differences can lead to disagreements. Significant personality differences between programmers and other roles within a project may cause clashes between roles. Understanding these personality differences may help game developers better navigate these conflicts and get along.  

The lower Extraversion seen in programmers may make communication between themselves and other roles (i.e., designers, artists and production members) less frequent. This can lead to communication breakdown, one example may be where design ideas, artwork or project scheduling that are beyond the capacity of programmers (perhaps due to time or resource constraints), and these issues are not communicated back to the originating teams resulting in plans incongruous with the product. Additionally, programmers' lower levels of Openness, contributing to being more close-minded and less welcoming to new experiences, may also cause clashes between the expectations and designs put forward by design teams when programmers are less willing to embrace novel ideas within the game-play or project setup. 

Being aware of these potential clashes should help members of multidisciplinary teams empathise better, and customize their development processes to overcome or mitigate some of these problems.

\subsubsection{Implications for Managers}

Management could introduce stress management techniques, and foster an awareness of crunch time in the industry (as well as introducing measures to mitigate it). If you are a programmer or employed in another role in this area you should also be made aware that crunch time could amplify existing traits such as Neuroticism. Additionally, programmers within the game development sphere should be conscious that when communicating with other teams, differences in Extraversion may lead to difficulties or misunderstandings during communication. Aiming to be more receptive to new ideas, acknowledging that software in the games industry (such as engine development) develops swiftly, and being aware of potential differences could mitigate issues---especially during crunch time.

Using these findings with a focus on workplace wellbeing, it can be seen as a way to help with conflict resolution. Personality scores could be used as a way to understand and aid in conflict resolution within the game development industry. For example, a designer with high levels of openness may suggest novel ideas for inclusion in the gameplay, but programmers scoring lower on openness may deem it too risky. This could result in disagreements between teams. Our research findings could help to provide context as to why these tensions arise. University courses could include information about personality differences seen across roles within game development to highlight these potential areas for conflict. In doing so, we are promoting understanding for the next generation of game developers.

Some literature ~\cite[e.g.][]{ebstrup2011association} suggests that less Neurotic employees have lower stress at work, and higher stress in the workplace has been linked to poor health outcomes~\cite{garfin2018acute}. Identifying those with predicting factors for burnout (e.g. high Neuroticism) could help employers pick individuals with, for example, higher Extraversion which has a negative correlation with burnout, but as Swider et al. note---an individual reporting high Extraversion may still burnout if put into a situation where (for example) lone working is expected on a regular basis~\cite{swider2010born}. Individual and workplace differences should therefore be examined together in any attempt to predict negative outcomes in the industry. 

General associations between tasks and personality traits do not imply that a person with different traits cannot excel in a task. However, some skills may be more difficult to acquire (e.g. introverts usually have to work harder at communication skills than extraverts). Understanding and acknowledging these potential differences can help managers develop better (case-by-case) strategies for overcoming personality-based conflicts. There is no universal approach because it depends on the problem and the specific people involved, but encouraging collaboration and emphasizing team cohesion probably help more often than not. 

\subsubsection{Implications for Recruiters and Career Counsellors}

For recruiters, knowing which personality traits are associated with which roles can help sort new employees into roles most appropriate for their personality to enhance group cohesion (similar to the methods described by Dounsga-Ard et al.~\cite{doungsa-ardSoftwareEngineeringPosition2020}). Similarly, Personality profiling is used in suggesting job roles, career paths and educational programs, especially for students who may have no specific career trajectory in mind yet.

However, this comes with issues over reductionism and privacy. Assuming that personality is the only factor for role suitability is inappropriate and overly simplistic as it discounts experience, knowledge, and multitudinous other factors. Furthermore, personality testing raises privacy issues: how much information should an employer hold about an employee and how should that information be used, especially if it can be construed as discrimination~\cite{smith2018personality}? 

Our finding could be used to assess role suitability, as part of a larger, more robust model, but should not be relied upon as a strong predictor of role success.

Finally, although personality is seen as an important part of selection processes, using personality assessment as a direct link to job performance is still controversial~\cite{nikolaou2018personnel}, so companies who invest in these metrics may not get concrete results, although it could be helpful as part of a wider range of assessments.

\subsection{Limitations}

Our results should be interpreted with caution, in light of several limitations.

Probability sampling is impractical for many reasons~\cite{baltes2020sampling}, not least of which is that no appropriate sampling frame of game development professionals exists, and we partially mitigated the problems with snowball sampling by using multiple recruitment methods (online and in-person). While we found no evidence of \textit{response bias}, there is no population-level data with which to assess \textit{sampling bias}. Our sample may not be representative of all the game developers in the UK, and is certainly not representative of all game developers worldwide. 

Comparing personality measures across studies (i.e., comparing game development professionals against non-game software development professionals) entails comparing basically convenience samples from different studies at different times, which is not as reliable as comparing subsets of a single sample, or data collected from a single study. We offer this comparison as part of the exploratory nature of the research, as well as providing a foundation for future work to build upon. 

Without access to the raw data from previous research, we cannot perform data diagnostics to assess the assumptions of our statistical tests. Furthermore, the scores derived for each sample were not necessarily from using the same IPIP items, and previous studies did not compute factor scores from items in the same way that we did. Both of these issues may reduce the robustness of our comparisons in ways we cannot measure. 

When dealing with multiple comparisons, such as in ANOVAs involving multiple groups, there is an increased chance for type I errors (falsely rejecting the null hypothesis and falsely detecting a statistically significant difference between groups). Therefore, these results should be interpreted conservatively~\cite{lee2018proper}.

We did not assess personality differences across age or gender because including too many analyses (e.g. age and gender splits) would increase the risk of type I errors~\cite{simmons2011false}. This may affect our results because personality varies with demographics, but not enough and not in the right directions to explain our results. For example, designers are more likely to be women than developers, and women tend to be more Neurotic~\cite{murphy2021international} and more Agreeable~\cite{weisberg2011gender} than men. However, we found that designers are more Extroverted and Open than developers. Similarly,  art is dominated by women, yet no personality differences were seen in our study. Nevertheless, demographic measures that were not measured or analysed could have contributed to the results.

\subsection{Future Work}

This exploratory work serves as the foundation for further pre-registered, confirmatory research, with sufficient statistical power to examine more interaction effects. A larger study, including participants from different areas (e.g. game development, app development, enterprise software) and roles (programming, art, production, product design, management), is needed to establish more firmly the personality differences we identify here. Such work is hindered by the lack of quality sampling frames \cite{baltes2020sampling}.

Beyond the need for a confirmatory replication, our results and the limitations of our study raise several important questions for future research:
\begin{enumerate}
    \item Why are certain roles associated with certain traits? Is it a selection process (e.g. more introverted people are drawn to programming), an effect of the work (e.g. programming makes people more neurotic), or some combination thereof? 
    \item Do personality traits associated with different roles actually predict performance or satisfaction in those roles? 
    \item To what extent do demographic and cultural differences intersect with the personality profiles of game developers? 
\end{enumerate}

Longitudinal research---i.e. following a cohort of game developers over several years, repeatedly testing their personality traits---is needed to address the first question. If working in the industry changes your personality, that has totally different implications from people being drawn to particular industries or roles because of their personality traits. 

Investigating the relationship between personality traits and job performance is easier said than done. While we have widely used, widely-validated, standards measures of personality, we do not have good measures of job performance. We don't know how to measure the overall success of software projects \cite{ralph2014dimensions}, or the productivity of individuals or teams \cite{sadowski2019rethinking}. There is no widely-accepted definition of job performance in game development or software engineering, or any role therein. Significant conceptual and instrumentation work is therefore needed to support any such project. 

Meanwhile, more research is needed to understand the intersection of culture and personality within games development. The next steps in this research would primarily be to collect a comparative sample for software engineers using the same methods and demographic population, and/or to collect a new set of data from game developers concurrently, to allow us to compare results directly between groups. 

In a somewhat different direction, various software practices (e.g. pair programming, mob programming, peer code review) and structures (e.g. remote work, multidisciplinary teams, team code ownership) probably work better for people with certain personality traits. We might hypothesize, for instance, that more extroverted developers enjoy pair programming more. More research is needed to understand the intersection between personality traits and software development practices. Such understanding could help managers and teams tailor their software development methods, understand breakdowns in otherwise effective practices, and overcome resistance to adopting effective practices.

\section{Conclusion}
\label{sec:Conclusion}
We conducted a questionnaire survey of 123 game development professionals based on the Five Factor Model~\cite{mccrae2008empirical}, using the 50-item IPIP scale~\cite{goldbergInternationalPersonalityItem2006}. After assessing construct validity, we examined differences between roles within game development, and differences between our sample and previous samples of software professionals and software engineering students. This analysis led to several novel contributions: 
\begin{enumerate}
    \item Game development professionals tend to have lower Agreeableness, Conscientiousness, Openness, and Extraversion, but higher Neuroticism than other software development professionals. 
    \item Game programmers tend to have lower Extraversion, and Openness, but higher Neuroticism, than their colleagues in art, design and production. 
\end{enumerate}

These results raise some very serious questions. Game development is thought to be among the most culturally toxic sub-fields of software engineering. Game development companies appear to have made little progress on key problems (e.g. crunch time and burnout) identified over 15 years ago~\cite{vanderhoef2015crunch}. Are more disagreeable, closed-minded, emotionally unstable people drawn to game development? Do more disagreeable, closed-minded, emotionally unstable people tend to succeed or at least stick with game development? Does working in game development make people more disagreeable, closed-minded and emotionally unstable? 

While this study cannot distinguish among these possibilities, all three appear highly problematic. Given the highly competitive, creative and pressured environment in which game developers work, identifying personality traits which might indicate success and resilience in such roles could have implications for industry employers. However, if being more disagreeable, closed-minded and neurotic makes employees \textit{more} resilient, something is very wrong. More research is plainly needed to determine what is going on in game development, and how long-term problems can be addressed.


\section*{Data Availability}
\label{sec:DataAvailability}

The supplementary materials, including data and analysis script are available at \url{osf.io/6gq2k/}. A Docker image for replication is also available, \url{hub.docker.com/r/ivorym/personalitygamedev-ease22}.

\balance


\end{document}